\documentclass[prd,amssymb,10pt]{revtex4}
\usepackage{amsmath,amsfonts,amssymb}
\usepackage{verbatim,graphicx,float}

\begin{document}

\title{Emerging singularities in the bouncing loop cosmology}

\author{Jakub Mielczarek}
\email{jakubm@poczta.onet.pl}
\affiliation{\it Astronomical Observatory, Jagiellonian University, 30-244
Krak\'ow, Orla 171, Poland}
\affiliation{\it The Niels Bohr Institute, Copenhagen University, Blegdamsvej 17, 
DK-2100 Copenhagen, Denmark}

\author{Marek Szyd{\l}owski}
\email{uoszydlo@cyf-kr.edu.pl}
\affiliation{Department of Theoretical Physics, 
Catholic University of Lublin, Al. Rac{\l}awickie 14, 20-950 Lublin, Poland}
\affiliation{Marc Kac Complex Systems Research Centre, Jagiellonian University,
Reymonta 4, 30-059 Krak{\'o}w, Poland}

%\date{\today}

\begin{abstract}
In this paper we calculate $\mathcal{O}(\mu^4)$ corrections from holonomies in 
the Loop Quantum Gravity, usually not taken into account. Allowance of the 
corrections of this kind is equivalent with the choice of the new quatization 
scheme. Quantization ambiguities in the Loop Quantum Cosmology allow for this 
additional freedom and presented corrections are consistent with the standard 
approach. We apply these corrections to the flat FRW cosmological model and 
calculate the modified Friedmann equation. We show that the bounce appears in 
the models with the standard $\mathcal{O}(\mu^2)$ quantization scheme is 
shifted to the higher energies $\rho_{\text{bounce}} = 3 \rho_{\text{c}}$. 
Also a pole in the Hubble parameter appears for $\rho_{\text{pole}} = 
\frac{3}{2} \rho_{\text{c}}$ corresponding to \emph{hyper-inflation/deflation} 
phases. This pole represents a curvature singularity at which the scale factor 
is finite. In this scenario the singularity and bounce co-exist. Moreover we 
find that an ordinary bouncing solution appears only when quantum corrections 
in the lowest order are considered. Higher order corrections can lead to the 
nonperturbative effects.
\end{abstract}

\maketitle

\section{Introduction} \label{sec:intro}

Strength of the gauge field $F$ in some point $x$ can be obtained form holonomy 
calculated around this point and taking limit of the zero length of the loop.
Loop Quantum Gravity (LQG) is kind of gauge theory describing gravitational 
degrees of freedom in terms of gauge field $A$ which is elements of 
$\mathfrak{su}(2)$ algebra and conjugated variable $E$ which is elements of 
$\mathfrak{su}(2)^*$ algebra \cite{Ashtekar:1987gu}. To quantise this theory 
in a background independent way one introduces holonomies of the Ashtekar 
connection $A$
\begin{equation}
h_{\alpha}[A] = \mathcal{P} \exp \int_{\alpha} A \quad \text{where 1-form} \quad A=\tau_i A^i_a dx^a
\label{holo}
\end{equation}
where $\tau_i = -\frac{i}{2}\sigma_i$ ($\sigma_i$ are Pauli matrices) and 
conjugated fluxes     
\begin{equation}
F_{S}^i[E] = \int_S dF^i \quad \text{where 2-form} \quad dF_i =\epsilon_{abc} E^a_i dx^b \wedge dx^c 
\end{equation}
as new fundamental variables \cite{Nicolai:2005mc,Perez:2004hj}. Other 
variables like the field strength $F$ should be expressed in term of these 
elementary variables. As we mentioned at the beginning the field strength can 
be expressed in term of holonomies. However, another aspect of loop 
quantisation starts to be important here. Namely, an area operator possesses 
a discrete spectrum with minimal nonzero eigenvalue $\Delta$ \cite{Ashtekar:1996eg}. 
So we cannot simply shrink to zero the area enclosed by loop.
Instead of this we must stop shrinking loop for a minimal value 
corresponding to the area gap $\Delta$. This effect leads to quantum gravitational 
corrections to the expression for classical field strength. The expression
for the field strength as a function of holonomies have a form \cite{Ashtekar:2006wn}
\begin{equation}
F^k_{ab} = \lim_{\mu \rightarrow \bar{\mu}} \left\{ - 2 
 \frac{\text{tr}\left[\tau_k \left( h^{(\mu)}_{\Box_{ij}}-\mathbb{I} \right)  \right]}{\mu^2 V_0^{2/3} } 
{^o\omega^i_a}{^o\omega^j_b}  +\frac{\mathcal{O}(\mu^4)}{\mu^2}\right\}
\label{lim}
\end{equation}
where the limit $\mu \rightarrow \bar{\mu}$ corresponds to the minimal value 
of the area gap $\Delta$. However, this formula is adequate only when 
$\mathcal{O}(\mu^4)$ terms can be neglected, i.e., in the classical limit.
In fact these terms, which form infinite series, are a function of $F$ and 
$A$.  The expression for the $F$ as a function of holonomies should be 
therefore obtained by solving this equation in terms of the first factor on 
the right side. In the classical limit $\mu \rightarrow 0$ terms 
$\mathcal{O}(\mu^4)/\mu^2$ vanish and we recover a classical expression for 
the field strength. Until now in literature the first order quantum correction 
to field strength has been investigated. It means that terms 
$\mathcal{O}(\mu^4)$ have been neglected. This approach was dictated be the 
choice of the simplest quantization scheme. Namely, as it has been shown by 
Bojowald \cite{Bojowald:2006gr,Bojowald:2007gc,Bojowald:2008pu}, the precise 
effective Hamiltonian must be a periodic function of the canonical variable 
$c$. The simplest form of this function we obtain when we perform the 
regularisation of the expression for the classical field strength cutting off 
the terms $\mathcal{O}(\mu^4)$. This is a standard procedure in the Loop 
Quantum Cosmology.

In this paper we calculate and study another non-vanishing contribution what 
is in fact a choice of the different regularisation of the expression for the 
field strength. It means that we hold $\mathcal{O}(\mu^4)$ factor, which is 
a function of $F$, and we solve equations for $F$ as a function of the 
holonomies. This approach is equivalent to the choice of the new quantization
scheme what is allowed due to quantization ambiguities.

The organisation of the text is the following. In section~\ref{sec:holo} 
we calculate expression for $F$ as a function of holonomies in 
$\mathcal{O}(\mu^4)$ order. Then in section~\ref{sec:FRW} we apply this result 
to the flat Friedmann-Robertson-Walker (FRW) cosmological model. We show that 
obtained correction have important influence for this model. In 
section~\ref{sec:summary} we summarise the results. Finally in the Appendix 
we give some basics of Loop Quantum Cosmology connected with the subject of 
this paper and explain the employed notation.

\section{Holonomy corrections} \label{sec:holo}

From the definition (\ref{holo}) we can calculate holonomy for
homogeneous model in the  particular direction $^oe^a_i\partial_a$ and
the length $\mu V_0^{1/3}$
\begin{equation}
h_{i}^{(\mu)} = e^{\tau_i \mu c } 
              = \mathbb{I}\cos \left( \frac{\mu c}{2}\right)+2\tau_i\sin \left( \frac{\mu c}{2}\right).
\label{hol2}
\end{equation}
From such particular holonomies we can construct a holonomy along the closed 
curve $\alpha=\Box_{ij}$ as schematically presented in the diagram below

\begin{center}
\setlength{\unitlength}{0.7cm}
\begin{picture}(10,10)
\thicklines
\put(3,3){\vector(1,0){2}}
\put(5,3){\line(1,0){2}}
\put(7,3){\vector(0,1){2}}
\put(7,5){\line(0,1){2}}
\put(7,7){\vector(-1,0){2}}
\put(5,7){\line(-1,0){2}}
 \put(3,7){\vector(0,-1){2}}
\put(3,5){\line(0,-1){2}}
\put(4.5,2){\vector(1,0){1}}
\put(5.5,8){\vector(-1,0){1}}
\put(8,4.5){\vector(0,1){1}}
\put(2,5.5){\vector(0,-1){1}}
\put(4.5,1.2){$^oe^a_i\partial_a$}
\put(8.3,5){$^oe^a_j\partial_a$}
\put(4.5,8.5){$-{^oe^a_i\partial_a}$}
\put(0.2,5){$-{^oe^a_j\partial_a}$}
\put(4.5,3.4){$h_{i}^{(\mu)}$}
\put(5.8,4.6){$h_{j}^{(\mu)}$}
\put(4.5,6){$h_{i}^{(\mu)-1}$}
\put(3.3,4.6){$h_{j}^{(\mu)-1}$}
\end{picture}
\end{center}

\noindent and can be written as
\begin{equation}
h_{\Box_{ij}}^{(\mu)} = h_{i}^{(\mu)} h_{j}^{(\mu)} h_{i}^{(\mu)-1} h_{j}^{(\mu)-1} 
= e^{\mu B_i}e^{\mu B_j}e^{-\mu B_i}e^{-\mu B_j}
\label{holo1}
\end{equation}
where we have introduced 
\begin{equation}
B_i := V_0^{1/3} A_a {^oe^a_i} = V_0^{1/3} c \tau_i. 
\label{def1}
\end{equation}
Factors $B_i$ are elements of $\mathfrak{su}(2)$ algebra so to perform product of exponents in equation~(\ref{holo1})
we need to use the Baker-Campbell-Hausdorff formula
\begin{eqnarray}
e^Xe^Y = \exp\left\{ X+Y+\frac{1}{2}[X,Y]+\frac{1}{12}\left([X,[X,Y]]+[Y,[Y,X]]\right) -\frac{1}{24}[Y,[X,[X,Y]]]
 +\dots   \right\}.
\end{eqnarray}
To calculate $\mathcal{O}(\mu^4)$ correction the elements of the expansion written above are sufficient.
Applying this formula to equation~(\ref{holo1}) we obtain
\begin{eqnarray}
h_{\Box_{ij}}^{(\mu)} &=&   \exp\left\{ \mu^2[B_i,B_j]+\frac{\mu^3}{2} [B_i+B_j,[B_i,B_j]] 
                      -\frac{\mu^4}{12}[B_j,[B_i,[B_i,B_j]]]+\frac{\mu^4}{6}[B_i+B_j,[B_i+B_j,[B_i,B_j]]]
 +\mathcal{O}(\mu^5)  \right\}     \nonumber \\
                      &=& \mathbb{I}+ \mu^2[B_i,B_j]+\frac{\mu^3}{2} [B_i+B_j,[B_i,B_j]] 
                      -\frac{\mu^4}{12}[B_j,[B_i,[B_i,B_j]]]+\frac{\mu^4}{6}[B_i+B_j,[B_i+B_j,[B_i,B_j]]]  \nonumber \\
                      &+& \frac{\mu^4}{2}[B_i,B_j][B_i,B_j]  +\mathcal{O}(\mu^5). 
\end{eqnarray} 
Now, multiplying this expression by $\tau_k$, using definition~(\ref{def1}) and taking a trace of 
both sides we obtain
\begin{eqnarray}
\text{tr}\left[\tau_k \left( h^{(\mu)}_{\Box_{ij}}-\mathbb{I} \right)  \right] &=&  
       \mu^2  c^2_{\text{h}}  \epsilon_{ijl}  \text{tr}\left( \tau_k \tau_l \right) + 
       \frac{\mu^3}{2} c^3_{\text{h}}  \epsilon_{ijl} (\epsilon_{ilm}+\epsilon_{jlm}) \text{tr}\left( \tau_k \tau_m \right)
-\frac{\mu^4}{12} c^4_{\text{h}} \epsilon_{ijl} \epsilon_{ilm} \epsilon_{jmn} \text{tr}\left( \tau_k \tau_n \right)\nonumber \\
 &+& \frac{\mu^4}{6} c^4_{\text{h}} \epsilon_{ijl} (\epsilon_{ilm}+\epsilon_{jlm}) (\epsilon_{imn}+\epsilon_{jmn}) 
\text{tr}\left( \tau_k \tau_n \right) + \frac{\mu^4}{2}c^4_{\text{h}} \epsilon_{ijl} \epsilon_{ijm}
  \text{tr}\left( \tau_k \tau_l \tau_m \right).
\label{equat9}
\end{eqnarray}
We mention that $\{ijk\}$ are external indices and the Einstein summation convention is not fulfilled.
The introduced parameter $c_{\text{h}}$ corresponds to the effective canonical variable $c$ which
is expressed as a function of holonomies. With use of equation~(\ref{hol2}) we can
directly calculate the left side of equation~(\ref{equat9}), we obtain 
\begin{equation}
\text{tr}\left[\tau_k \left( h^{(\mu)}_{\Box_{ij}}-\mathbb{I} \right)  \right] =
 - \frac{\epsilon_{kij}}{2} \sin^2\left(\mu c \right).
\label{tr}
\end{equation}
Then, using properties of $\tau_i$ matrices we obtain
\begin{equation}
\frac{1}{3}\mu^4 c^4_{\text{h}} - \mu^2 c^2_{\text{h}} + \sin^2\left(\mu c \right) = 0.
\end{equation}
The $\mathcal{O}(\mu^3)$ order contribution simply vanishes.
The solutions of this equation have a form
\begin{equation}
c^2_{\text{h}\pm}  = \frac{1\pm\sqrt{1-\frac{4}{3}\sin^2\left(\mu c \right) }}{\frac{2}{3}\mu^2}.
\end{equation}
When we expand the square in the solution for $c^2_{\text{h}-}$ we obtain 
\begin{equation}
c^2_{\text{h}-} = \left[ \frac{\sin (\mu c)}{\mu}  \right]^2 +\frac{1}{3}\frac{\sin^4 (\mu c)}{\mu^2} +\dots. 
\end{equation}
The first factor of the expansion corresponds to the known case when $\mathcal{O}(\mu^4)$ corrections
are ignored. We can easily check than the classical limit $\mu \rightarrow 0,\ c^2_{\text{h}}\rightarrow c^2$ is
recovered only in the $c^2_{\text{h}-}$ case. The case $c^2_{\text{h}+}$ should be therefore 
treated as unphysical. However, as we will see in the next section, both solutions 
lead to the same modified Friedmann equation. So we can keep both solutions.

Finally the expression for the effective field strength has a form
\begin{equation}
F_{ab}^k = \epsilon^k_{\ ij}\frac{1\pm\sqrt{1-\frac{4}{3}\sin^2\left(\bar{\mu} c \right) }}{\frac{2}{3}\bar{\mu}^2 V_0^{2/3}} 
{^o\omega^i_a}{^o\omega^j_b}.
\label{EffFS}
\end{equation}
We have performed here the limit $\mu \rightarrow \bar{\mu}$ where 
\begin{equation}
\bar{\mu} = \sqrt{\frac{\Delta}{|p|}}.
\label{mubar}
\end{equation}
For details of this limit see papers \cite{Ashtekar:2006wn} or appendices to 
the papers \cite{Chiou:2007mg,Mielczarek:2008zv,Magueijo:2007wf}.

As we have mentioned earlier the precise effective Hamiltonian must be 
a periodic function of $c$. In our case the effective Hamiltonian has
a form $H_{\text{eff}} \sim \sqrt{|p|} c^2_{\text{h}-} $ where
the  $c^2_{\text{h}-}$  can be expressed as  
\begin{equation}
 c^2_{\text{h}-} = \frac{1}{2\bar{\mu}^2} \sum_{n=1}^{\infty} \frac{(2n)!}{(2n-1)n!^2 3^{n-1}(2i)^{2n}} 
\left[ \exp(i\bar{\mu} c)-\exp(-i\bar{\mu} c) \right]^{2n}.
\end{equation}
As we see this function is periodic, and forms an infinite series numerated by 
integers. However, this infinity is allowed in the frames of Loop Quantum 
Cosmology. The obtained effective Hamiltonian is correct however is not given 
by the simple function as we should expect for fundamental expressions.
However, we should to keep in mind that we are looking for the effective 
Hamiltonian and there is no circumstances that such a Hamiltonian must 
have mathematically simple allowed form.

In the next section we will use the calculated effective field strength $F$
for the FRW $k=0$ cosmological model.

\section{Application to FRW $k=0$} \label{sec:FRW}

\begin{figure}[ht!]
\centering
$\begin{array}{cc}   
\includegraphics[width=6cm,angle=270]{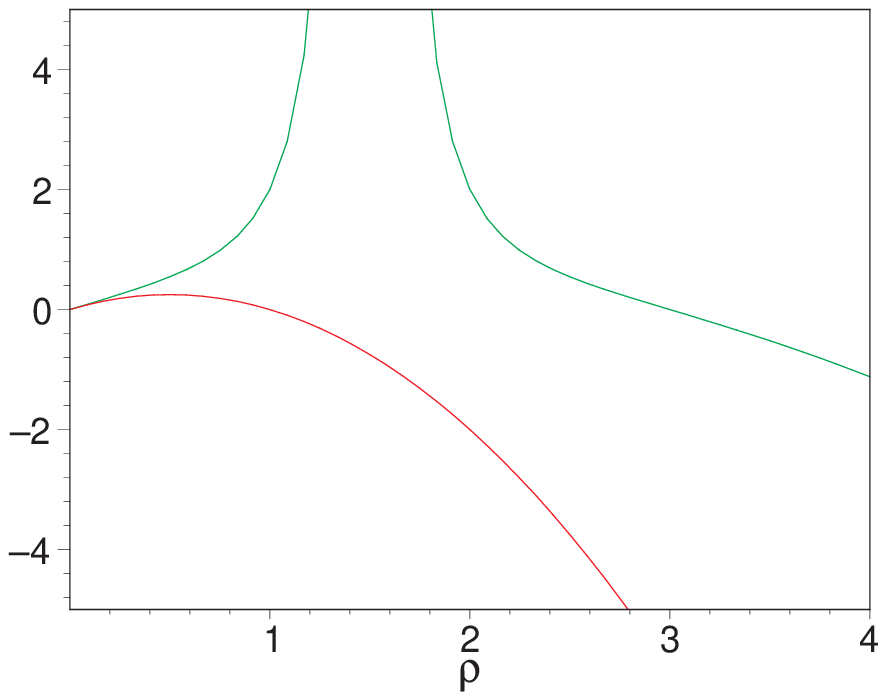}  &  \includegraphics[width=6cm,angle=270]{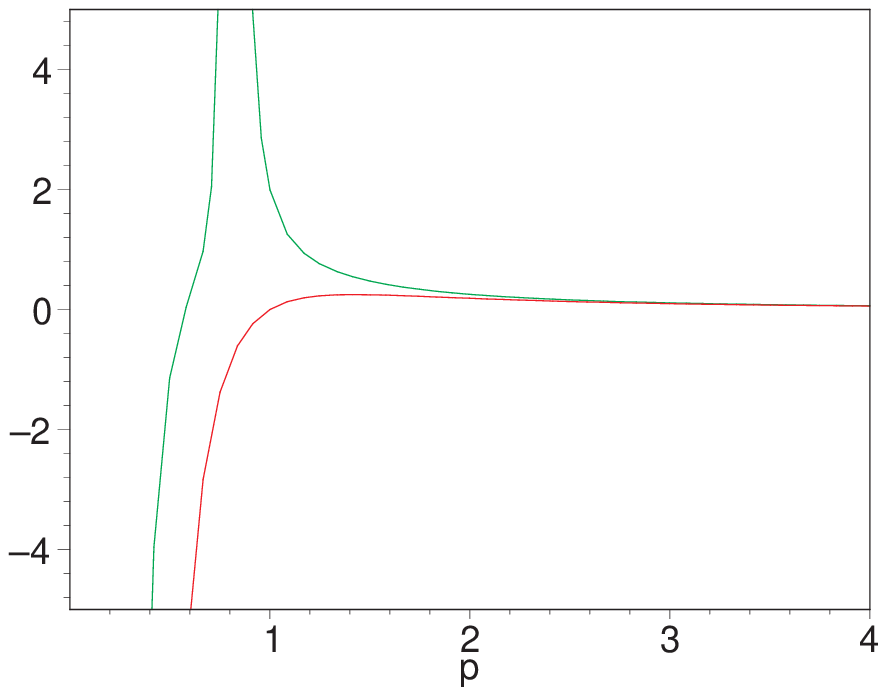}   
\end{array}
$\caption{ 
 {\bf Left }: Evolution of $H^2_{\mathcal{O}(\mu^2) }$ (red, bottom curve)  and $H^2_{\mathcal{O}(\mu^4) }$ 
 (green, top curve) as a function of energy density $\rho$. The parts below zero value of $H^2$ are unphysical.
The $\rho$ is in the units of $\rho_{\text{c}}$. 
 { \bf Right }: Evolution of $H^2_{\mathcal{O}(\mu^2) }$ (red, bottom curve)  and $H^2_{\mathcal{O}(\mu^4) }$ 
 (green, top curve) as a function of $p$ for the model with radiation $(\rho \propto 1/p^2)$. 
On the both panels a physically admissible region corresponds to $H^2\geq 0$. }
\label{HubblePlots}
\end{figure}

With use of equation (\ref{EffFS}) we can derive the effective Hamiltonian for 
the flat FRW model in the form
\begin{equation}
H_{\text{eff}} =  - \frac{3}{8 \pi G \gamma^2}
 \frac{1\pm\sqrt{1-\frac{4}{3}\sin^2\left(\bar{\mu} c \right) }}{\frac{2}{3}\bar{\mu}^2}  \sqrt{|p|}  +{|p|}^{3/2}\rho.
\label{model}
\end{equation}
For details we send to the appendix. This Hamiltonian fulfils the so called 
Hamiltonian constraint $H_{\text{eff}} = 0$. From Hamilton equations we can 
calculate evolution of the canonical variable $p$
\begin{equation}
\dot{p} = \left\{p,H_{\text{eff}}  \right\}  = - \frac{8\pi G \gamma}{3} \frac{\partial H_{\text{eff}} }{\partial c}
\end{equation}
and with use of (\ref{model}) we obtain 
\begin{equation}
\dot{p} = \mp \frac{\sqrt{|p|}}{\gamma \bar{\mu}}
\frac{2 \sin(\bar{\mu}c) \cos(\bar{\mu}c)}{\sqrt{1-\frac{4}{3}\sin^2\left(\bar{\mu} c \right) } }. 
\label{dotp}
\end{equation}
Applying equation (\ref{dotp}), the Hamiltonian constraint $H_{\text{eff}} = 0$ and definition 
of the Hubble parameter $H= \frac{\dot{p}}{2p}$ we finally derive the
modified Friedmann equation
\begin{equation}
H^2_{\mathcal{O}(\mu^4) }= \frac{8\pi G}{3}\rho \left(1-\frac{\rho}{3\rho_{\text{c}}}  \right)
\left[\frac{3}{4}+\frac{1}{4}\frac{1}{\left(1-\frac{2}{3}\frac{\rho}{\rho_{\text{c}}}  \right)^2 }  \right]
\label{Fried1}
\end{equation}
where we have introduced 
\begin{equation}
\rho_{\text{c}} =  \frac{\sqrt{3}}{16\pi^2 \gamma^3 l_{\text{Pl}}^4}.
\end{equation} 
As we see obtained equation does not depend on the sign $\pm$ in the Hamiltonian. 
An analogous equation in the lowest order has been calculated earlier \cite{Ashtekar:2006wn}
and have a form
\begin{equation}
H^2_{\mathcal{O}(\mu^2) }= \frac{8 \pi G}{3}\rho \left(1-\frac{\rho}{\rho_{\text{c}}}  \right).
\end{equation}
This equation lead to the bounce for $\rho=\rho_{\text{c}}$.
An analogous bounce is also present in the derived model (\ref{Fried1}), however now the bounce is shifted
to the higher energy densities
\begin{equation}
\rho_{\text{bounce}} = 3 \rho_{\text{c}}.
\end{equation}
Another important property is the pole in the Hubble parameter for 
\begin{equation}
\rho_{\text{pole}} = \frac{3}{2} \rho_{\text{c}}
\end{equation}
as we see from equation (\ref{Fried1}). We show these features in Fig.~\ref{HubblePlots}.
In the left panel we present $H^2$ as a function of energy density for $\mathcal{O}(\mu^2)$ 
and  $\mathcal{O}(\mu^4)$ cases. In the right panel we compare evolution of the 
Hubble parameter as a function of $p$ for the radiation dominated Universe $(\rho \propto 1/p^2)$.

In both case we observe a nonperturbative feature, namely the pole for 
$\rho = \frac{3}{2} \rho_{\text{c}}$.
This fact indicates that higher order corrections from holonomies can have important 
influence for dynamical behaviour for small values of $p$. It is clear 
from a parameter of expansion (\ref{mubar}) which grows for small values of $p$.
For large $p$ the classical case is clearly recovered, however behaviour
for small values of $p$ is highly complicated. Namely, as our 
study suggests higher order terms of expansion have nonperturbative 
influence for dynamics and this fact can seriously complicate 
a simple bouncing universe picture. We investigate this 
issue in the next section.

\section{Qualitative analysis of dynamics} \label{sec:}

The advantage of qualitative methods of analysis of differential equations 
\cite{Perko} is that we obtain all evolutional paths for all admissible 
initial conditions. In this approach the evolution of the system is represented 
by trajectories in the phase space and asymptotic states by critical points. 
We demonstrate that dynamics of the model can be reduced to 2-dimensional 
autonomous dynamical system. These methods allow to distinguish a generic 
evolutional scenario.

\begin{figure}[ht!]
\centering
\includegraphics[width=6cm,angle=0]{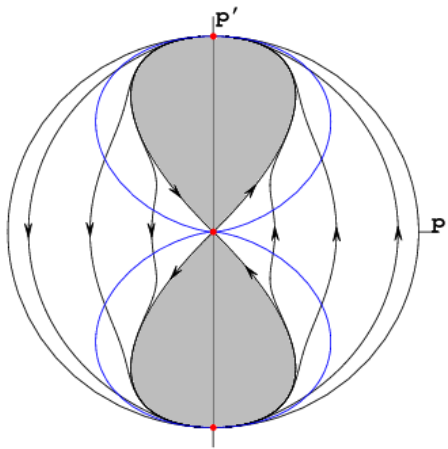}
\caption{The phase portrait for all admissible initial conditions. 
The blue line is representing points at which trajectories pass horizontally 
through the inflection point (\emph{hyper-inflationary/deflationary} phases).}
\label{fig:3}
\end{figure}

In Fig.~\ref{fig:3} we show the phase portrait for all admissible initial 
conditions (all values of total energy $E$ of the fictitious particle moving 
in a $1$-dimensional potential proportional to $p^3$) for the model with the 
free scalar field.

\begin{figure}[ht!]
\centering
\includegraphics[width=10cm,angle=0]{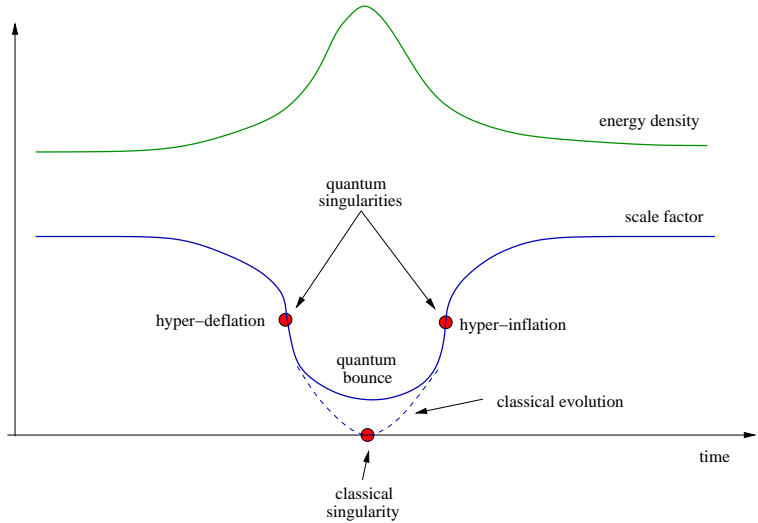}
\caption{The schematic picture of the evolution of the model. The top curve 
(green) represents the energy density of the free scalar field. It is worth to 
note that this energy density is finite during whole evolution, even during 
transitions through singularities. The bottom curve (blue) presents schematic 
evolution of the scale factor for the investigated model. The dashed curve 
represents the classical evolution which is not realised in the presented 
quantum model.}
\label{fig:4}
\end{figure}

The physical trajectories are situated in the region at which $E-V$ is 
non-negative or $p$ is larger than some minimal value and $p'$ is zero. The 
physical trajectories lie in the non-shaded region bounded by a zero velocity 
curve which represents a homoclinic orbit. Of course the whole system is 
symmetric with respect to the reflection ($H$ is changed in $-H$). The blue 
line on the phase portrait represents points at which trajectories pass 
horizontally through the inflection point (\emph{hyper-inflationary/deflationary} 
phases). Therefore, evolution comprises the bounce solution interpolating 
static phases of evolution, see Fig. \ref{fig:4}. There is the intermediate 
phase of evolution at which we have the inflection point at the diagram of 
$p(t)$. Note that it is the singularity state with rapid growth of the scale 
factor, we call this phase the hyper-inflation.
It is important to note that energy density is finite during transition
through these singularities. Similar finite scale factor singularities 
has been studied recently by Cannata~\emph{et al.\/} \cite{Cannata:2008xc}.

In the phase diagram in Fig.~\ref{fig:3} we adjoint a circle at infinity in
standard way via Poincare construction. Note that all trajectories are starting from 
unstable node - representing static Einstein Universe and landing at stable node representing
another static Einstein Universe.

Parisi \emph{et al.\/} \cite{Parisi:2007kv} pointed out that stability of Einstein static models
in high-energy modifications of General Relativity is important from the point of view of 
so-called Emergent Universe scenario \cite{Ellis:2002we}. 
Mulryne \emph{et al.\/} \cite{Mulryne:2005ef} investigated the
stability of the Einstein static model and they found that LQC Einstein static model is
representing a centre type of critical point on the phase portrait. As it is well known such
a critical point is structurally unstable. Note that in our case this static universe
is representing a node type of critical point. This modification of stability in presented
model has important consequences for the Emergent Universe scenario, since as it is well
known in General Relativity, a static Universe is unstable and is represented by a saddle type 
of critical point and therefore it requires the fine-tuning. Moreover corrections considered
lead to a regularization of the big-bang singularity \cite{Bojowald:2001xe}.

\section{Summary} \label{sec:summary}

In this paper we have calculated another non-vanishing contribution to the quantum
holonomy correction in Loop Quantum Gravity. Quantum correction of this kind 
appears when we express the Ashtekar connection $A$ and field strength $F$ in terms
of holonomies. The source of quantum modification to classical expressions
is a non-vanishing area enclosed by loop as the result of existence of the area gap $\Delta$.

We had applied obtained corrections to the flat FRW cosmological model 
and the we have calculated resulting quantum gravitational modifications
to the Friedmann equation. The holonomy correction in the 
lowest order to the flat FRW model has been calculated earlier
\cite{Ashtekar:2006rx,Ashtekar:2006uz,Ashtekar:2006wn}
and extensively studied \cite{Singh:2006im, Mielczarek:2008zv}. These investigations uncovered existence 
of the bounce for energy scales $\rho_{\text{c}}$.
In this picture the standard Big Bang singularity is replaced by the non-singular
Big Bounce. Calculations performed in the present paper indicate that the
holonomy correction in the next non-vanishing order holds this picture.
Namely, the initial singularity is still preserved. However the bounce appears
now for higher energy density  $\rho_{\text{bounce}} = 3 \rho_{\text{c}}$.
Another important new feature is the appearance of a pole in the 
Hubble parameter for $\rho_{\text{pole}} = \frac{3}{2} \rho_{\text{c}}$
corresponding to \emph{hyper-inflationary/deflationary\/} phases.
This leads to more complicated dynamical behaviour at these energy scales.

We showed that the generic evolutional scenario for the model with
the free scalar field starts from the static 
Einstein universe then recolapse passing through the curvature singularity 
with a finite scale factor (hyper-deflation) towards the bounce and goes in 
the expanding phase through the second curvature singularity (hyper-inflation)
and ends in the static Einstein 
universe. During the transition through the singularities the universal 
critical behaviour $H \propto | \rho - \rho_{\text{pole}} |^{-1}$ holds. 
Therefore in the presented scenario the bounce connects these two finite scale 
factor singularities. 

As we see, the higher order quantum correction in LQG can have an important 
influence on dynamical behaviour of cosmological models. It is not unlikely 
that the non-singular bounce appeared in the lowest order can be only an 
artifact of simplifications and can disappear when whole contribution will be 
taken into account. Further investigations of quantum corrections from LQG are 
still necessary. We conclude that higher order holonomy corrections 
and resulting different quantization schemes should be also seriously 
taken into account in considerations.

\begin{acknowledgments}
We thank to Martin Bojowald for useful comments and prof. Jerzy 
Lewandowski and {\L}ukasz Szulc for discussion during the workshop ``Quantum 
Gravity in Cracow'' 12-13.01.2008. 
Authors are grateful to Tomasz Stachowiak for stimulating discussion. 
This work was supported in part by the Marie Curie Actions Transfer of
Knowledge project COCOS (contract MTKD-CT-2004-517186) and project 
Particle Physics and Cosmology (contract MTKD-CT-2005-029466).
\end{acknowledgments}

\appendix 
\section{Flat FRW model in Loop Quantum Gravity} \label{Appendix1}

The FRW $k=0$ spacetime metric can be written as
\begin{equation}
ds^2=-N^2(x) dt^2 + q_{ab}dx^adx^b
\end{equation}
where $N(x)$ is the lapse function and the spatial part of the metric is expressed as 
\begin{equation}
q_{ab}= \delta_{ij} {\omega^i_a} {\omega^j_b}= a^2(t) {^oq}_{ab} = a^2(t)  \delta_{ij}  {^o\omega^i_a}{^o\omega^j_b}.
\end{equation}
In this expression ${^oq}_{ab}$ is fiducial metric and ${^o\omega^i_a}$ are co-triads dual to the triads 
${^oe^a_i}$,  ${^o\omega^i}({^oe_j})=\delta^i_j$   where $^o\omega^i={^o\omega^i_a}dx^a$ and $^oe_i={^oe_i^a}\partial_a$.
From these triads we construct the Ashtekar variables 
\begin{eqnarray}
A^i_a &\equiv& \Gamma^i_a+\gamma K_a^i = c V_0^{-1/3} \ {^o\omega^i_a}  , \label{A} \\
E^a_i &\equiv& \sqrt{|\det q|} e^{a}_i = p V_0^{-2/3} \sqrt{^oq} \ {^oe^a_i} \label{E}
\end{eqnarray}  
where 
\begin{eqnarray}
|p| &=& a^2 V_0^{2/3}, \\
 c &=& \gamma \dot{a} V_0^{1/3}.
\end{eqnarray}
Note that the Gaussian constraint implies that $p \leftrightarrow -p$
leads to the same physical results.
The factor $\gamma$ is called the Barbero-Immirzi parameter, $\gamma=\ln 2 / (\pi \sqrt{3} )$.
In the definition (\ref{A}) the spin connection is defined as
\begin{equation}
\Gamma^i_a = -\epsilon^{ijk}e^b_j(\partial_{[a}e^k_{b]}+\frac{1}{2}e^c_k e^l_a \partial_{[c}e^l_{b]} )
\end{equation}
and the extrinsic curvature is defined as 
\begin{equation}
K_{ab}=\frac{1}{2N}\left[ \dot{q}_{ab} -2 D_{(a}N_{b)} \right]
\end{equation}
what corresponds to $K^i_a := K_{ab} e^b_i$.

The scalar constraint, in the Ashtekar variables, has the form 
\begin{eqnarray}
H_{\rm G} = \frac{1}{16 \pi G} \int_{\Sigma} d^3 x N(x) \frac{E^a_i
E^b_j}{\sqrt{|\mathrm{det} E|}}  \left[  {\varepsilon^{ij}}_k F_{ab}^k-2(1+\gamma^2)  K^i_{[a}  K^j_{b]} \right]
\label{scalar}
\end{eqnarray}
where field strength is expressed as
\begin{equation}
F^k_{ab}=\partial_aA^k_b-\partial_bA^k_a+\epsilon^k_{\ ij}A^i_aA^j_b.
\label{str}
\end{equation}
With use of  (\ref{A}),(\ref{E}) and (\ref{str}) the Hamiltonian
(\ref{scalar}) assumes the form
\begin{equation}
H_{\text{G}} = - \frac{3}{8 \pi G \gamma^2} \sqrt{|p|} c^2
\label{hami}
\end{equation}
where we have assumed a gauge of $N(x)=1$. Quantum corrections to this Hamiltonian come
when we express $\sqrt{|p|}$ and $c^2$ in terms of background independent variables. 
In this paper we had concentrated on the corrections to the factor $c^2$, 
called holonomy corrections. For a short review of quantum corrections we 
send to the appendix in the paper \cite{Magueijo:2007wf}.

\end{document}